\documentclass{elsart}
\usepackage{color,setspace,times}
\usepackage{amssymb,amsmath,graphicx,color,rotating,subfigure,url}
\usepackage[square,sort&compress,comma]{natbib}
\usepackage{lineno}

\journal{Physica A}

\begin{document}

\begin{frontmatter}

\title{Empirical regularities of order placement in the Chinese stock market}
\author[SB,SS]{Gao-Feng Gu},
\author[SZSE]{Wei Chen},
\author[SB,SS,RCE,RCSE]{Wei-Xing Zhou\corauthref{cor}}
\corauth[cor]{Corresponding address: 130 Meilong Road, P.O. Box 114,
School of Business, East China University of Science and Technology,
Shanghai 200237, China. Tel.: +86 21 64253634; fax: +86 21
64253152.}
\ead{wxzhou@ecust.edu.cn} %

\address[SB]{School of Business, East China University of Science and Technology, Shanghai 200237, P. R. China}
\address[SS]{School of Science, East China University of Science and Technology, Shanghai 200237, P. R. China}
\address[SZSE]{Shenzhen Stock Exchange, 5045 Shennan East Road, Shenzhen 518010, P. R. China}
\address[RCE]{Research Center for Econophysics, East China University of Science and Technology, Shanghai 200237, P. R. China}
\address[RCSE]{Research Center of Systems Engineering, East China University of Science and Technology, Shanghai 200237, P. R. China}

\begin{abstract}
Using ultra-high-frequency data extracted from the order flows of 23
stocks traded on the Shenzhen Stock Exchange, we study the empirical
regularities of order placement in the opening call auction, cool
period and continuous auction. The distributions of relative
logarithmic prices against reference prices in the three time
periods are qualitatively the same with quantitative discrepancies.
The order placement behavior is asymmetric between buyers and
sellers and between the inside-the-book orders and outside-the-book
orders. In addition, the conditional distributions of relative
prices in the continuous auction are independent of the bid-ask
spread and volatility. These findings are crucial to build an
empirical behavioral microscopic model based on order flows for
Chinese stocks.
\end{abstract}


\end{frontmatter}

\section{Introduction}

Stylized facts are common statistical characters observed from
different stocks within different time periods and usually presented
in a qualitative form. Concerning equity returns, the stylized facts
contain the absence of autocorrelations, heavy tails,
multifractality and intermittency, volatility clustering, leverage
effect, and so on
\cite{Mantegna-Stanley-2000,Cont-2001-QF,Zhou-2007}. Many stylized
facts in financial markets can be reproduced with microscopic
models. If a microscopic model presents mock stylized facts in
conformity with real ones, this model is believed to have catched
some underlying regularities of financial markets. Based on how the
price forms, there are two types of microscopic models for financial
markets known as the agent-based models and order driven models
\cite{Zhou-2007}. The price variations in agent-based models are
determined by the imbalance between demand and supply, including
percolation models
\cite{Cont-Bouchaud-2000-MeD,Stauffer-1998-AP,Stauffer-Penna-1998-PA,Eguiluz-Zimmermann-2000-PRL,DHulst-Rodgers-2000-IJTAF,Xie-Wang-Quan-Yang-Hui-2002-PRE},
Ising models
\cite{Foellmer-1974-JMe,Chowdhury-Stauffer-1999-EPJB,Iori-1999-IJMPC,Kaizoji-2000-PA,Bornholdt-2001-IJMPC,Zhou-Sornette-2007-EPJB},
minority games
\cite{Arthur-1994-AER,Challet-Zhang-1997-PA,Challet-Marsili-Zhang-2000-PA,Jefferies-Hart-Hui-Johnson-2001-EPJB,Challet-Marsili-Zhang-2001-QF,Challet-Marsili-Zhang-2001a-PA,Challet-Marsili-Zhang-2001b-PA,Challet-Marsili-Zhang-2005},
and others
\cite{Bak-Paczuski-Shubik-1997-PA,Lux-Marchesi-1999-Nature}. The
price in order-driven models changes based on the continuous double
auction (CDA) mechanism
\cite{Maslov-2000-PA,Mike-Farmer-2008-JEDC,Daniels-Farmer-Gillemot-Iori-Smith-2003-PRL,Farmer-Patelli-Zovko-2005-PNAS}.
Two fundamental ingredients of order-driven models are order
placement and order cancelation \cite{Mike-Farmer-2008-JEDC}. The
regularities governing the dynamical processes of order placement
and cancelation can be determined empirically in some sense. In this
way, very realistic behavioral models can be constructed.

Order placement plays a key role in the simulation of price
formation in order driven models, since the proportion of placed
orders is much greater than that of canceled orders. When placing an
order, the trader need to determine its sign (``$+1$'' for buys and
``$-1$'' for sells), size and price. In determining the order price,
the trader faces a situation of dilemma and has to balance two
contradictive factors, the certainty of execution on one hand and
the potential benefit on the other hand. Patient traders possibly
consider the fact of benefit more important than the other and place
orders inside the limit-order book with a less aggressive price
(high price for sellers and low for buyers). The situation is
different for impatient traders who consider the factor of execution
certainty in the first place. This kind traders want to make a
transaction as soon as possible and place the order outside the
limit-order book with a more aggressive price (low for sellers and
high for buyers). In this work, we focus on the prices of submitted
orders.

Zovko and Farmer studied the unconditional distribution of relative
limit prices defined as the distance from the same best prices for
orders placed inside the limit-order book
\cite{Zovko-Farmer-2002-QF}. They merged the data from 50 stocks
traded on the London Stock Exchange (August 1, 1998 to April 31,
2000) and found that the distribution decays roughly as a power law
with the tail exponent approximately $\alpha = 1.5$ for both buy and
sell orders. Bouchaud {\em et al}. analyzed the order books of three
liquid stocks on the Paris Bourse (February 2001) and found that the
relative price of new orders placed inside the book follows a
power-law distribution with the tail exponent $\alpha = 0.6$
\cite{Bouchaud-Mezard-Potters-2002-QF}. Potters and Bouchaud
investigated the relative limit price distributions for
inside-the-book orders of three Nasdaq stocks (June 1 to July 15,
2002) and found that the distributions exhibit power-law tails with
an exponent $\alpha = 1$ \cite{Potters-Bouchaud-2003-PA}. Maskawa
analyzed 13 rebuild order books of Stock Exchange Electronic Trading
Service from July to December in 2004 on the London Stock Exchange
and found that the limit prices for all orders inside the book are
broadly distributed with a power-law tail whose exponent is $\alpha
= 1.5$ \cite{Maskawa-2007-PA}, which is consistent with the results
of Zovko and Farmer \cite{Zovko-Farmer-2002-QF}. He also presented
the distribution in the negative part for more aggressive order
outside the book and found that the negative part decays much faster
than the positive part. Mike and Farmer focused on the stock named
AZN and tested on 24 other stocks listed on the London Stock
Exchange \cite{Mike-Farmer-2008-JEDC}. They found that the
distribution of relative logarithmic prices can be fitted by a
Student distribution with $\alpha=1.0-1.65$ degrees of freedom and
the distribution is independent of bid-ask spread at least over a
restricted range for both buy and sell orders.

There are also efforts to seek for factors influencing order
placement. Using 15 stocks on the Swiss Stock Exchange, Ranaldo
found that both bid-ask spread and volatility negatively relate to
order aggressiveness \cite{Ranaldo-2004-JFM}. Lillo analyzed the
origin of power-law distribution of limit order prices with the
method of considering the order placement as an utility maximization
problem considering three factors: time horizon, utility function
and volatility \cite{Lillo-2007-EPJB}. He found that the
heterogeneity in time horizon is the proximate cause of the
power-law distribution, while heterogeneity in volatility is hardly
connected with the origin of power-law distribution.

The paper is organized as follows. In Section~\ref{s1:dataset}, we
explain the data set analyzed and briefly introduce the trading
rules of the Shenzhen Stock Exchange (SSE). Section~\ref{s1:UD}
presents the unconditional probability distributions of relative
prices in three periods: opening call auction, cool period, and
continuous double auction. We then study in Section~\ref{s1:CD} the
conditional probability distributions against bid-ask spread and
volatility, respectively. The last section concludes.

\section{Dataset}
\label{s1:dataset}

We analyze a huge database containing the order flows of 23 liquid
stocks listed on the Shenzhen Stock Exchange in the whole year 2003
\cite{Gu-Chen-Zhou-2008-PA}. The order flow records
ultra-high-frequency data whose time stamps are accurate to 0.01
second including details of every event. Each limit order can be
identified by the order placement time. The logarithmic price of an
order at time $t$ is denoted as $\pi(t)$. The tick size of the
quotation price of an order is RMB 0.01 yuan. As an emerging stock
market, with the purpose of speculation limitation and healthy
development, the Exchange imposes a daily price limit of 10\% on
trading of stocks, which means that the maximum price fluctuation on
trading day $T$ must be restricted to ten percent of the closing
price $p(T-1)$ of the previous trading day. More details about the
trading rules can be found in Ref. \cite{Gu-Chen-Zhou-2007-EPJB}.

The Chinese stock market is an order-driven market with the key
mechanism called continuous double auction (CDA). Before July 1,
2006, only limit orders are allowed. There are three time periods
for open call auction (9:15 a.m. to 9:25 a.m.), cool period (9:25
a.m. to 9:30 a.m.), and continuous double auction (9:30 a.m. to
11:30 a.m. and 13:00 p.m. to 15:00 p.m.). Table~\ref{Tb:NOP} shows
the numbers of valid buy and sell orders placed in the three
periods.

\begin{table}[htp]
 \centering
 \caption{An overview of order placement}
 \medskip
 \label{Tb:NOP}
 \centering
 \begin{tabular}{lrrrrrrrrr}
 \hline \hline
    && \multicolumn{8}{@{\extracolsep\fill}c}{Number of orders} \\
    \cline{3-10}
    Stock code&
    & \multicolumn{2}{@{\extracolsep\fill}c}{Opening call auction} &
    & \multicolumn{2}{@{\extracolsep\fill}c}{Cool period} &
    & \multicolumn{2}{@{\extracolsep\fill}c}{Continuous auction} \\
    \cline{3-4} \cline{6-7} \cline{9-10}
    &
    & buy~~ & sell~~ &
    & buy~~ & sell~~ &
    & buy~~~~ & sell~~~~ \\
    \hline
    000001 && 45,719 & 72,685 && 21,495 & 22,905 && 1,650,942 & 1,500,371 \\
    000002 && 24,098 & 48,296 && 8,898 & 13,017 && 826,806 & 891,475 \\
    000009 && 19,766 & 41,028 && 9,365 & 14,518 && 838,207 & 904,612 \\
    000012 && 8,368 & 18,192 && 5,882 & 6,490 && 463,536 & 437,421 \\
    000016 && 7,276 & 14,568 && 4,045 & 4,526 && 286,046 & 308,869 \\
    000021 && 13,239 & 24,387 && 9,265 & 9,935 && 660,818 & 695,660 \\
    000024 && 5,640 & 12,631 && 2,867 & 4,284 && 191,098 & 232,446 \\
    000027 && 13,435 & 35,007 && 5,142 & 9,123 && 467,385 & 670,321 \\
    000063 && 10,394 & 23,800 && 4,055 & 6,139 && 354,840 & 466,066 \\
    000066 && 9,532 & 19,860 && 6,272 & 6,768 && 453,067 & 489,532 \\
    000088 && 3,092 & 8,645 && 1,462 & 2,096 && 121,927 & 153,971 \\
    000089 && 9,519 & 19,313 && 3,650 & 4,863 && 297,322 & 303,752 \\
    000406 && 15,315 & 31,300 && 7,287 & 10,011 && 462,696 & 457,572 \\
    000429 && 7,045 & 13,505 && 2,888 & 3,369 && 184,443 & 207,910 \\
    000488 && 9,095 & 15,104 && 2,888 & 3,795 && 179,374 & 183,640 \\
    000539 && 5,030 & 13,718 && 2,291 & 3,945 && 179,984 & 170,520 \\
    000541 && 7,034 & 12,936 && 1,701 & 2,530 && 105,055 & 119,808 \\
    000550 && 9,936 & 20,427 && 5,828 & 7,504 && 521,182 & 566,326 \\
    000581 && 5,115 & 13,531 && 2,455 & 3,032 && 146,821 & 181,723 \\
    000625 && 12,516 & 23,481 && 6,425 & 8,375 && 555,846 & 582,814 \\
    000709 && 13,324 & 27,200 && 5,709 & 7,558 && 351,883 & 375,383 \\
    000720 && 9,536 & 16,433 && 1,382 & 2,673 && 145,797 & 133,175 \\
    000778 && 8,858 & 22,771 && 3,224 & 6,193 && 236,609 & 288,828 \\
    \hline \hline
 \end{tabular}
\end{table}

\section{Unconditional distributions of relative logarithmic prices}
\label{s1:UD}

We define the relative price $x$ as the logarithmic distance of
order price from a reference price, which is presented as follows,
\begin{equation}
x\left(t\right) = \left\{
 \begin{array}{ccl}
 \pi(t) - \pi_{r_1}(t-1) && ~~{\rm{for~buy~orders}} \\
 \pi_{r_2}(t-1) - \pi(t) && ~~{\rm{for~sell~orders}}
 \end{array}
 \right.,
\label{Eq:x}
\end{equation}
where $\pi(t)$ is the logarithmic price of a coming order at time
$t$, $\pi_{r_1}(t-1)$ and $\pi_{r_2}(t-1)$ are the logarithmic
reference prices right before the order is placed at time $t$.
Roughly speaking, orders with larger relative prices $x$ are more
aggressive.

In the Chinese stock market, there is a daily price limit of 10\%.
If the closure price of a stock on a trading day is $p$, then the
price of any new order in the successive trading day is constrained
in the range $[p_{\min},p_{\max}]=[R(0.9p),R(1.1p)]$, where $R(y)$
is the round number of $y$. The definition domain of the relative
price $x$ is
\begin{equation}
[\pi_{\min},\pi_{\max}]=[\ln(0.9/1.1),\ln(1.1/0.9)] \approx
[-0.2007,0.2007]~.
\end{equation}
Therefore, all the distribution diagrams in this work have fixed
abscissa width. In addition, no matter what the functional form of
the distribution is, arbitrary moments of $x$ exist.

In the ensuing subsections, we study the probability distributions
of relative prices $x$ for 23 stocks in the three periods (opening
call auction, cool period and continuous auction) with different
trading mechanisms. We have compared the distributions for
individual stocks and found that they are remarkably analogous. This
is not surprising since the same traders invest in different stocks
which makes the order placement dynamics underlying different stocks
behave similarly. We thus treat all the stocks as an ensemble and
study the distribution aggregating all the 23-stock data in each
period. This treatment is also supported by the fact that $x$ is
dimensionless and varies within $[-0.2007,0.2007]$ for all stocks.

\subsection{Distributions in opening call auction}

Opening call auction is held between 9:15 and 9:25 on each trading
day. It refers to the process of one-time centralized matching of
buy and sell orders accepted during this time period. At any time
$t$, the virtual transaction price $\pi_v(t)$ is determined
according to the following principles: (i) the price that generates
the greatest trading volume; (ii) the price which allows all the buy
orders with higher bid price and all the sell orders with lower
offer price to be executed; (iii) the price which allows either buy
side or sell side to have all the orders identical to such price to
be executed. Orders are executed together at the end of opening call
auction. The word ``virtual'' means that the orders are not really
executed and the price is disposed to all traders. Whenever a new
order arrives or an order is canceled\footnote{According to the
trading rules, no cancelation is allowed between 9:20 a.m. and 9:25
a.m. in the opening call auction process.}, the virtual price is
updated. However, we focus on new prices only when new orders are
submitted, which are stamped with time $t$. In other words, the time
$t$ increases by one step after a new valid order is placed.

\begin{figure}[htb]
\centering
\includegraphics[width=6.5cm]{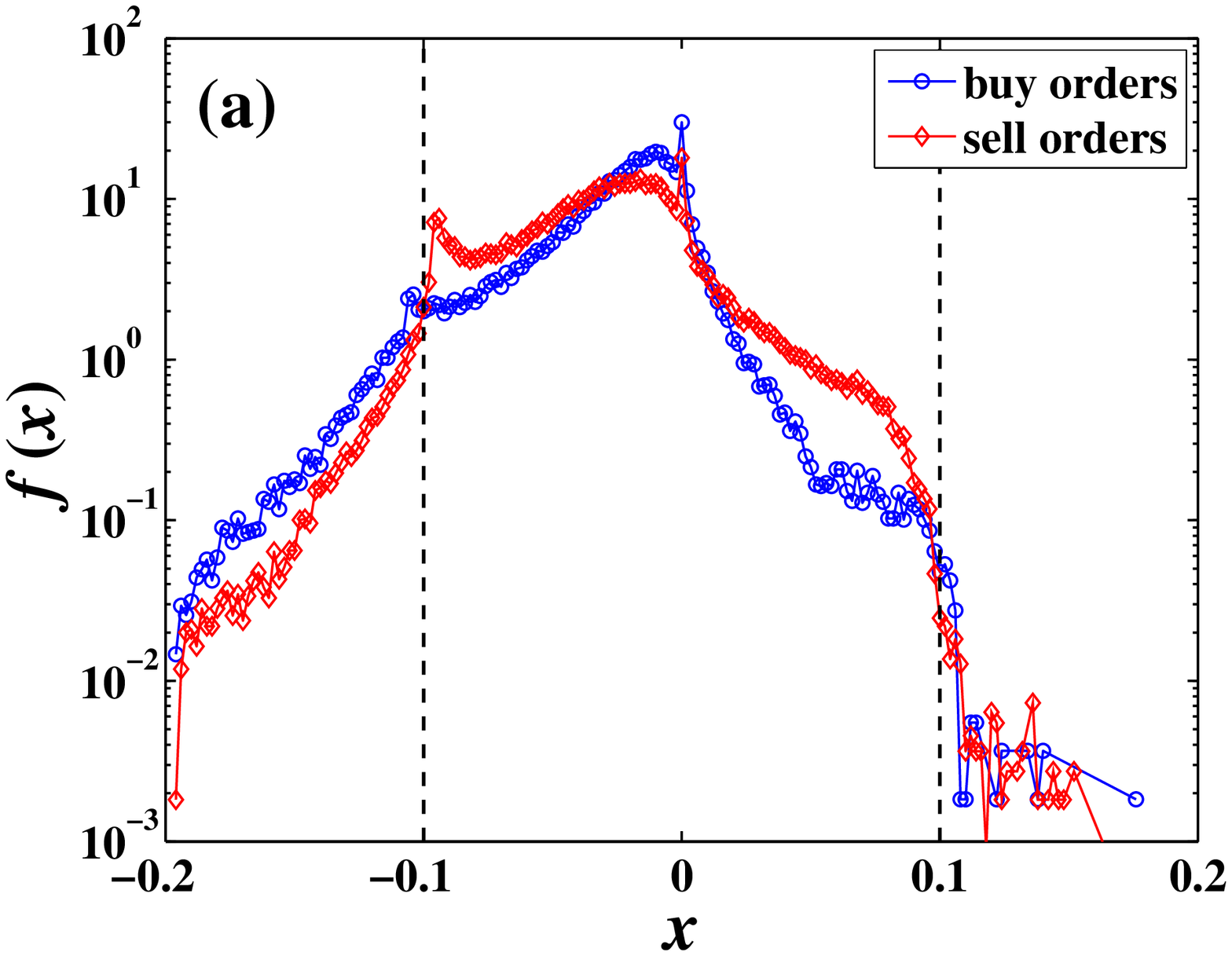}
\includegraphics[width=6.5cm]{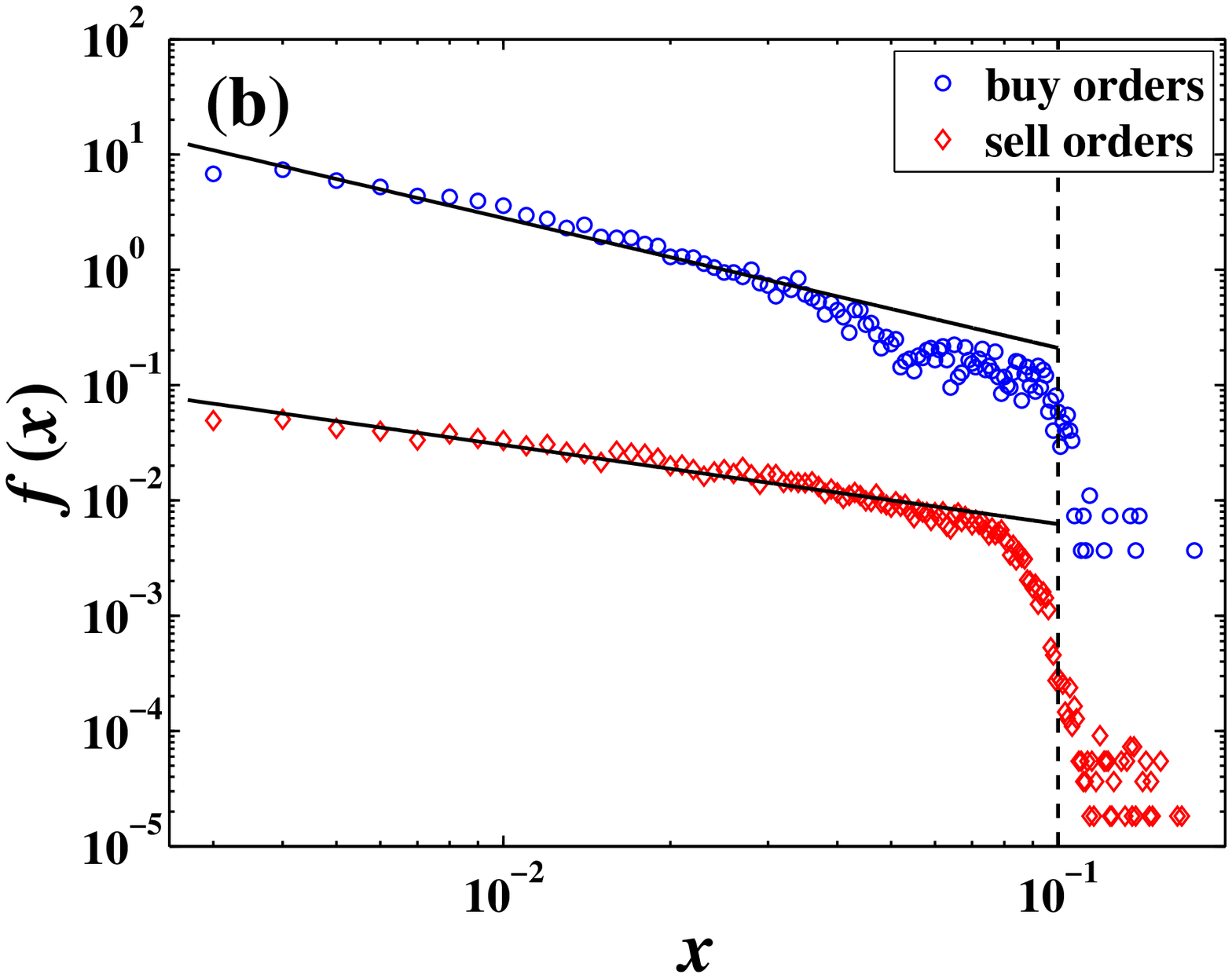}
\caption{\label{Fig:PDF_OCA_BS}(Color online) Panel (a): Empirical
probability density functions $f(x)$ of relative prices $x$
aggregating the 23-stock data in the opening call auction for buy
orders and sell orders, respectively. Panel (b): Log-log plot of the
probability density functions $f(x)$ for the positive relative
prices $x > 0$. The solid lines are the least squares fits of
Eq.~(\ref{Eq:OCA:fx:PL}) to the data with $\alpha_{\rm{buy}} = 0.13
\pm 0.04$ for buy orders and $\alpha_{\rm{sell}} = -0.31\pm 0.02$
for sell orders. The plot for sell orders has been vertically
translated downward for clarity.}
\end{figure}

According to the definition of relative price $x$ in
Eq.~(\ref{Eq:x}), in the opening call auction we have
\begin{equation}
 \pi_{r_1}(t) = \pi_{r_2}(t) = \pi_v(t-1),
\label{Eq:oca}
\end{equation}
where $\pi_v(t-1)$ is the virtual transaction price at time $t-1$.
Fig.~\ref{Fig:PDF_OCA_BS} presents the probability distribution of
relative prices $x$ for buy orders (circles) and sell orders
(diamonds) in the opening call auction.

Fig.~\ref{Fig:PDF_OCA_BS}(a) shows that there are three local maxima
in the density function $f(x)$ for both kinds of orders. Each $f(x)$
function reaches the global maximum value when $x = 0$, which means
that the virtual price $\pi_v$ plays an important role in the
process of order placement. A large proportion (3.46\%) of the
orders are placed on the virtual price to increase the execution
probability. The secondary maximum locates at negative $x$ close to
$x=0$, which means that the traders are more patient: buyers pose
higher price while seller put lower price in order to lower their
costs with satisfying certainty of execution. The third maximum is
close to $x=10\%$, the daily price limit. In conformity with the
trading rule of daily price limit, a large number of patient traders
place their orders at the lowest price (for buyers) or the highest
price (for sellers) no matter what the current price is. On the
other hand, the probability distribution of overnight returns
reaches it maximum at zero. The third maximum is thus explained
based on these two facts. There is also another significant kink
around $x=0.1$, which is also induced by the price limit rule.

The distribution of $x$ is asymmetric between the orders placed
inside the order book ($x < 0$) and those outside the book ($x >
0$). The density $f(x)$ decays more slowly for orders inside the
book. Speaking differently, more traders consider to curtail the
cost of investment and are less aggressive. In addition, the
distribution for buy orders are not the same as that of sell orders.
There are more buy orders placed close to the virtual price. The
underlying mechanism for this asymmetry between buy and sell orders
are not clear.

In Fig.~\ref{Fig:PDF_OCA_BS}(b), we plot the probability density
$f(x)$ as a function of relative price for outside-the-book orders
on log-log coordinates. We find that each distribution roughly
follows a power law in the bulk
\begin{equation}
 f(x) \sim x^{-(1+\alpha)}~.
 \label{Eq:OCA:fx:PL}
\end{equation}
Linear least-squares fitting gives the power-law exponents
$\alpha_{\rm{buy}} = 0.13 \pm 0.04$ for buy orders in the range
$0.002 \leqslant x \leqslant 0.045$ and $\alpha_{\rm{sell}} =
-0.31\pm 0.02$ for sell orders in the range $0.002 \leqslant x
\leqslant 0.072$.

\subsection{Distributions in cool period}

Following the opening call auction, cool period begins at 9:25 and
ends at 9:30. Within this period, the trading system is open to
orders routing from members, but does not process orders or
cancelation. The trading system presents the information at the
first three best on each side. We term it ``cool period'', because
the information including prices and volumes released by the
Exchange and displayed on the terminal screens does not change
within this time period. Traders place their orders according to
those fixed reference information.

\begin{figure}[htb]
\centering
\includegraphics[width=6.5cm]{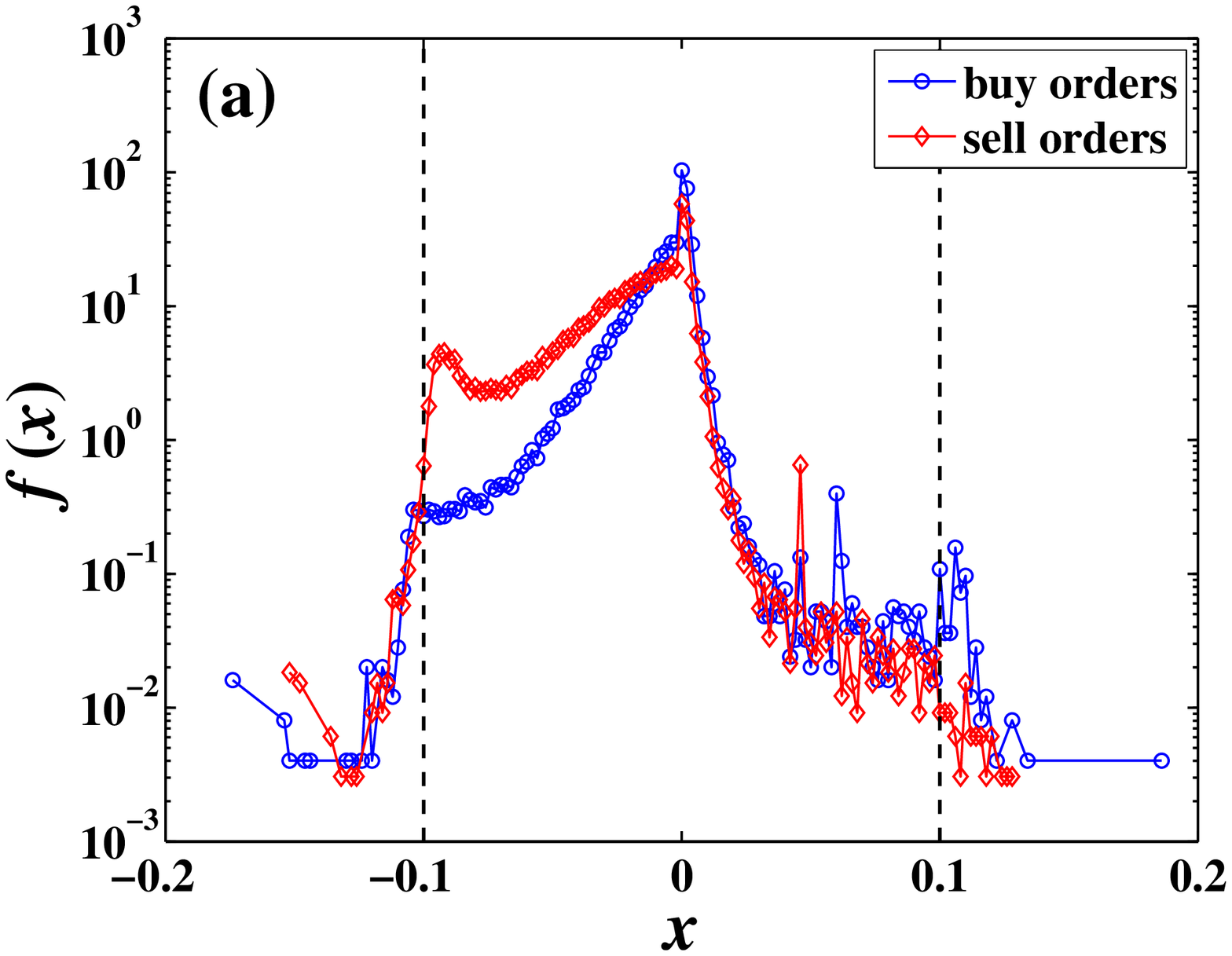}
\includegraphics[width=6.5cm]{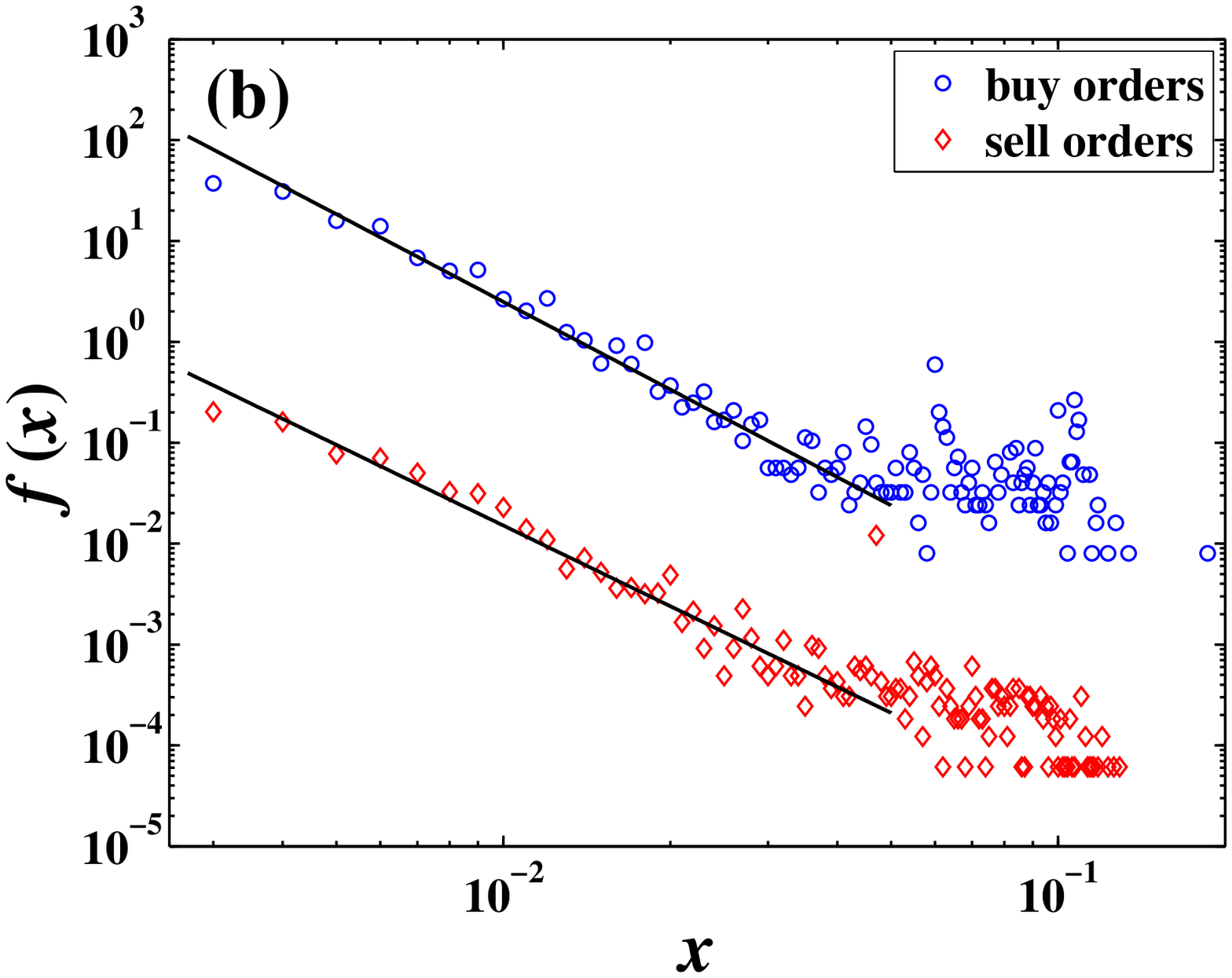}
\caption{\label{Fig:PDF_CL_BS}(Color online) Panel (a): Empirical
probability density functions $f(x)$ of relative prices $x$
aggregating the 23-stock data in the cool period for buy orders and
sell orders, respectively. Panel (b): Log-log plot of the
probability density functions $f(x)$ for the positive relative
prices $x > 0$. The solid lines are the least squares fits of
Eq.~(\ref{Eq:OCA:fx:PL}) to the data with $\alpha_{\rm{buy}} = 1.89
\pm 0.08$ for buy orders and $\alpha_{\rm{sell}} = 1.66 \pm 0.09$
for sell orders. The plot for sell orders has been vertically
translated downward for clarity.}
\end{figure}

The relative price $x$ of a new order is defined as the logarithmic
distance from the same best price\footnote{The same best price is
the best bid price for buy orders and the best ask price for sell
orders.}. In the definition of relative price $x$ presented in
Eq.~(\ref{Eq:x}), the reference prices are
\begin{equation}
 \left\{
 \begin{array}{ccc}
 \pi_{r_1}(t) &=& \pi_b\\
 \pi_{r_2}(t) &=& \pi_a
 \end{array}
 \right.~,
\label{Eq:cl}
\end{equation}
where $\pi_b$ and $\pi_a$ are respectively the best bid and best ask
which keep constant and are independent of time $t$. The probability
density functions $f(x)$ of relative prices $x$ for buy orders and
sell orders in the cool period are shown in
Fig.~\ref{Fig:PDF_CL_BS}.

As we can see in Fig.~\ref{Fig:PDF_CL_BS}(a), the distributions are
similar to their counterparts in the opening call auction
qualitatively. However, the probability at $x = 0$ is higher than
that in the opening call auction and the second maximum in the
opening call auction disappears. Since there are much less data
points in the cool period than in the open call auction, the $f(x)$
functions have large fluctuations in this time period.

In Fig.~\ref{Fig:PDF_CL_BS}(b), we present the distributions of the
positive relative prices $x > 0$ on log-log coordinates. The main
bodies follow a power-law behavior with the exponent
$\alpha_{\rm{buy}} = 1.89 \pm 0.08$ in the range $0.003 \leqslant x
\leqslant 0.04$ for buy orders and $\alpha_{\rm{sell}} = 1.66 \pm
0.09$ in the range $0.003 \leqslant x \leqslant 0.04$ for sell
orders.

\subsection{Distributions in continuous auction}

Continuous auction (9:30 a.m. - 11:30 a.m. and 13:00 p.m. - 15:00
p.m.) is the main part of the trading process. It refers to the
process of continuous matching of buy orders and sell orders on a
one-by-one basis. The orders not executed during the opening call
auction and placed in the cool period automatically enter the
continuous auction. The execution price in the continuous auction
can be determined according to the following principles: (i) where
the highest bid price matches the lowest offer price, the deal is
concluded at such price; (ii) where the bid price is higher than the
lowest offer price currently available in the central order book,
the deal is concluded at the lowest offer price; (iii) where the
offer price is lower than the highest bid price currently available
in the central order book, the deal is concluded at the highest bid
price. The trading system shows the three best prices and their
volumes by the interaction of the limit-order book and order flow.
Note that there is no closing call auction (14:57 p.m. - 15:00 p.m.)
in the year 2003, which was effective since July 1 2006 to generate
the closing prices.

\begin{figure}[htb]
\centering
\includegraphics[width=6.5cm]{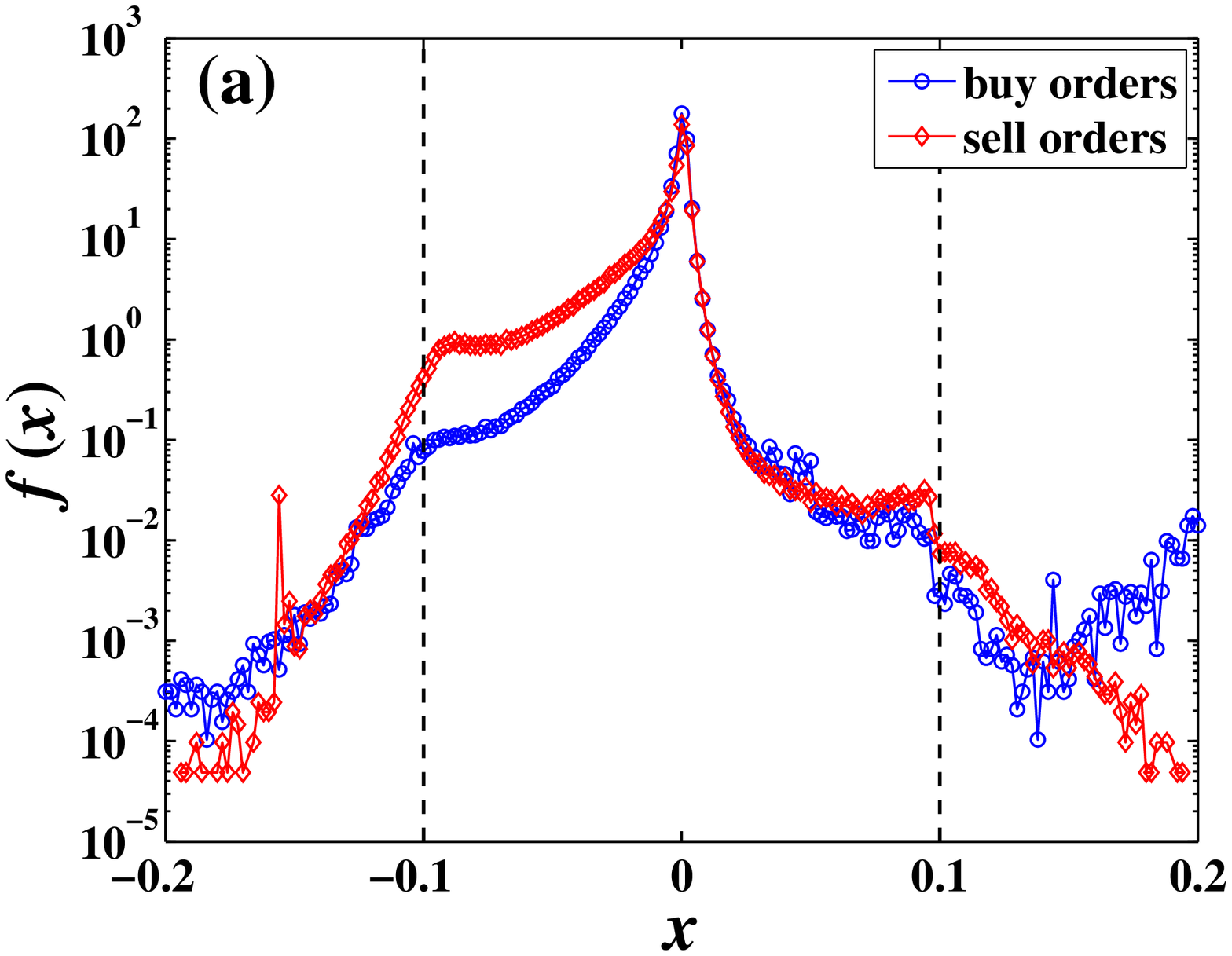}
\includegraphics[width=6.5cm]{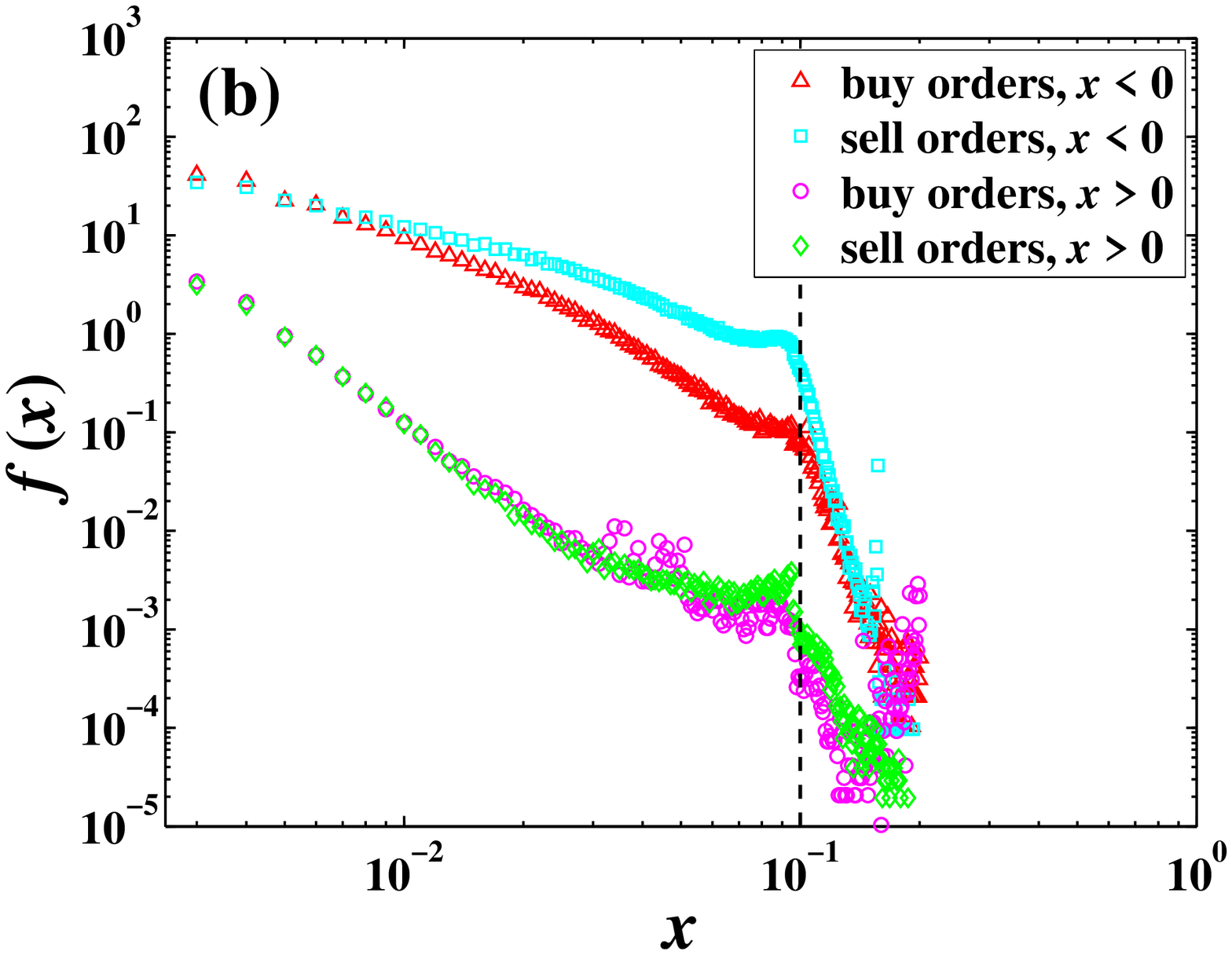}
\caption{\label{Fig:PDF_CDA_BS}(Color online) Panel (a): Empirical
probability density functions $f(x)$ of relative prices $x$
aggregating the 23-stock data in the continous auction for buy
orders and sell orders, respectively. Panel (b): Log-log plot of the
probability density functions $f(x)$. The plots for outside-the-book
orders has been vertically translated downward for clarity.}
\end{figure}

The two reference prices in the definition of relative price $x$
presented in Eq.~(\ref{Eq:x}) are chosen as the same best prices:
\begin{equation}
 \left\{
 \begin{array}{ccc}
 \pi_{r_1}(t) &=& \pi_b(t)\\
 \pi_{r_2}(t) &=& \pi_a(t)
 \end{array}
 \right.~,
\label{Eq:CDA}
\end{equation}
where $\pi_b(t)$ and $\pi_a(t)$ are the best bid and best ask in the
continuous auction which are variable with respect to time $t$. The
probability density functions $f(x)$ of relative prices $x$ for buy
orders and sell orders are shown in Fig.~\ref{Fig:PDF_CDA_BS}.

The two distributions for Chinese stocks illustrated in
Fig.~\ref{Fig:PDF_CDA_BS}(a) exhibit very different behavior when
compared with other stock markets
\cite{Zovko-Farmer-2002-QF,Bouchaud-Mezard-Potters-2002-QF,Potters-Bouchaud-2003-PA,Maskawa-2007-PA,Mike-Farmer-2008-JEDC}.
The most import idiosyncratic feature is that there exist kinks
around $x=\pm0.1$, which is induced by the 10\% price limit trading
rule in the Chinese market. The second discrepancy concerns the
asymmetry between buy orders and sell orders. The distributions for
both types of orders are identical in the interval $x\in[0,0.1]$,
beyond which they deviate significantly. In the interval
$x\in[-0.1,0]$, the density $f(x)$ is greater for sell orders than
for buy orders indicating that sellers are more patient to place
orders inside the book with less aggressive prices. In contrast,
there is no significant difference between the distributions for buy
orders and sell orders of stocks on the London Stock Exchange
\cite{Maskawa-2007-PA,Mike-Farmer-2008-JEDC}. We also find that the
distribution for each types of orders in the Chinese market is
asymmetric with respect to the same best price ($x=0$): there are
more orders placed inside the book. This is qualitatively consistent
with the result of Maskawa \cite{Maskawa-2007-PA} but different from
that of Mike and Farmer \cite{Mike-Farmer-2008-JEDC}.

When $|x|>0.1$, three tails of the distributions decay
exponentially. Compared with Fig.~\ref{Fig:PDF_CDA_BS}(b), these
tails look like power laws. These two descriptions can be unified
since $\ln|x|= |x|-1$ for $|x|-1$ being not too greater than 0. For
aggressive buy orders with $x>0.1$, the tail distribution exhibits
an abnormal upward trend. Since there were only limit orders allowed
for placement, the best strategy for an trader who wants to have her
order executed immediately is to place the order at the highest
price allowed $\pi_{\max}$ to buy or the lowest price allowed
$\pi_{\min}$ to sell. This consideration partly explain this
anomaly. However, it is not clear why the anomaly exists only for
buy orders. According to this strategy, one expects to see possible
local maxima around $x=0.1$ for both buy and sell orders, which is
actually the case. For patient traders, the 10\% price limit rule
provides a simple strategy to place orders at the lowest price
$\pi_{\min}$ to buy or the highest price $\pi_{\max}$ to sell. This
strategy applies especially to those traders who want to ``catch the
bottom'' or ``escape the roof''\footnote{``Catch the bottom'' is a
terminology frequently used by Chinese traders when the market is
experiencing a severe drop for several successive days. After
several days of price drop, more and more traders will try to
predict the end of the correction or crash and place buy orders at
the projected bottom. On the contrary, if the market is bullish for
a while and more and more people think that it is a bubble, traders
might take the strategy to sell before the market reverses. It is
termed as escaping the roof.}. This strategy also applied to
short-term speculators who think that the intraday price may
fluctuate largely\footnote{In order to reduce the market risks and
speculation actions, the Chinese stock market adopts $T+1$ trading
system, which does not allow traders to sell the stocks bought on
the same day. However, $T+0$ trading strategy is frequently adopted
by speculators. If a confident speculator project that the price of
a held stock will fluctuate significantly, she can buy new shares at
lower price and sell held shares at higher price. The order of
buying and selling is determined by her projection of the intraday
price trend.}.

In Fig.~\ref{Fig:PDF_CDA_BS}(b), we plot the two sides of each
density function on double logarithmic coordinates. Each curve
decays as a power law in a certain scaling range. For the positive
relative prices, we obtain $\alpha_{\rm{buy}}^{+} = 1.66 \pm 0.07$
in the range $0.003 \leqslant x \leqslant 0.04$ for the buy orders
and $\alpha_{\rm{sell}}^{+} = 1.80 \pm 0.06$ in the range $0.003
\leqslant x \leqslant 0.04$ for the sell orders. These power-law
exponents are greater than those of the London Stock Exchange
\cite{Maskawa-2007-PA,Mike-Farmer-2008-JEDC}. When we focus on the
distributions of negative relative prices, we have
$\alpha_{\rm{buy}}^{-} = 1.72 \pm 0.03$ in the range $0.003
\leqslant x \leqslant 0.04$ for buy orders and
$\alpha_{\rm{sell}}^{-} = 1.15 \pm 0.02$ in the range $0.003
\leqslant x \leqslant 0.05$ for sell orders. These exponents are
comparable to the results discovered on the London stock Exchange to
some extent
\cite{Zovko-Farmer-2002-QF,Mike-Farmer-2008-JEDC,Maskawa-2007-PA}.
However, all these exponents are greater than those discovered on
the Paris Bourse and Nasdaq stock market
\cite{Bouchaud-Mezard-Potters-2002-QF,Potters-Bouchaud-2003-PA}.

\section{Conditional distribution}
\label{s1:CD}

When placing orders, traders may consider other factors, such as the
bid-ask spread, volatility, limit-order depth, and so on. Here, we
focus on one of the most liquid stock (Shenzhen Development Bank Co.
Ltd, code 000001, see also Table \ref{Tb:NOP}) in the continuous
auction and check whether the probability density function of
relative prices is dependent on the bid-ask spread and volatility,
respectively.

\subsection{Conditional distribution on spread}
\label{s2:CD:spd}

Bid-ask spread, defined as the difference between the best ask price
and the best bid price, is considered as the benchmark of the
transaction cost and a measure of market liquidity. The definition
of bid-ask spread in literature varies and we adopt the definition
put forward by Mike and Farmer \cite{Mike-Farmer-2008-JEDC}
\begin{equation}
 s(t) = \ln\pi_a(t) - \ln\pi_b(t)~,
\label{Eq:st}
\end{equation}
whose statistical properties has been studied for the Chinese stocks
\cite{Gu-Chen-Zhou-2007-EPJB}.

\begin{figure}[htb]
\centering
\includegraphics[width=6.5cm]{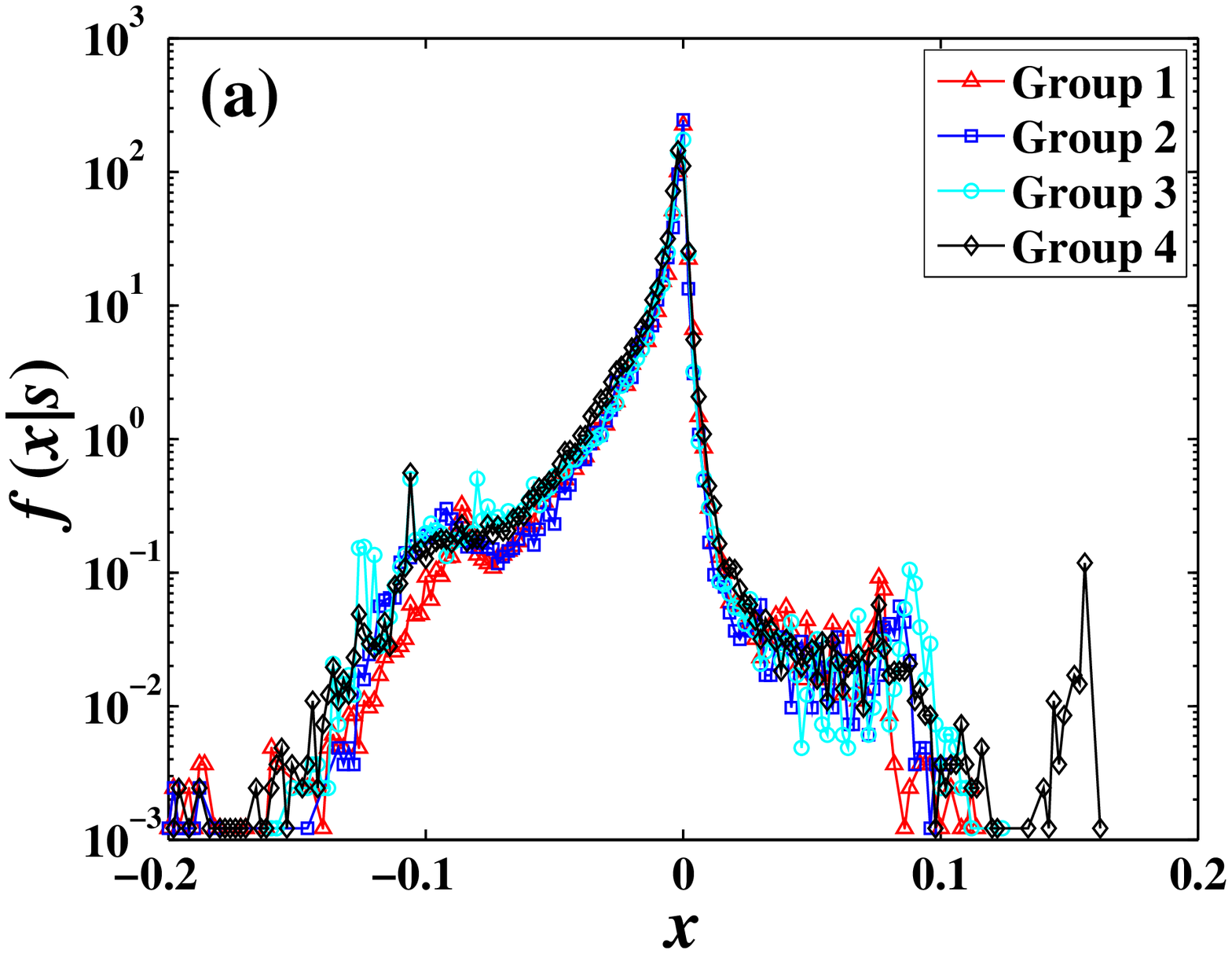}
\includegraphics[width=6.5cm]{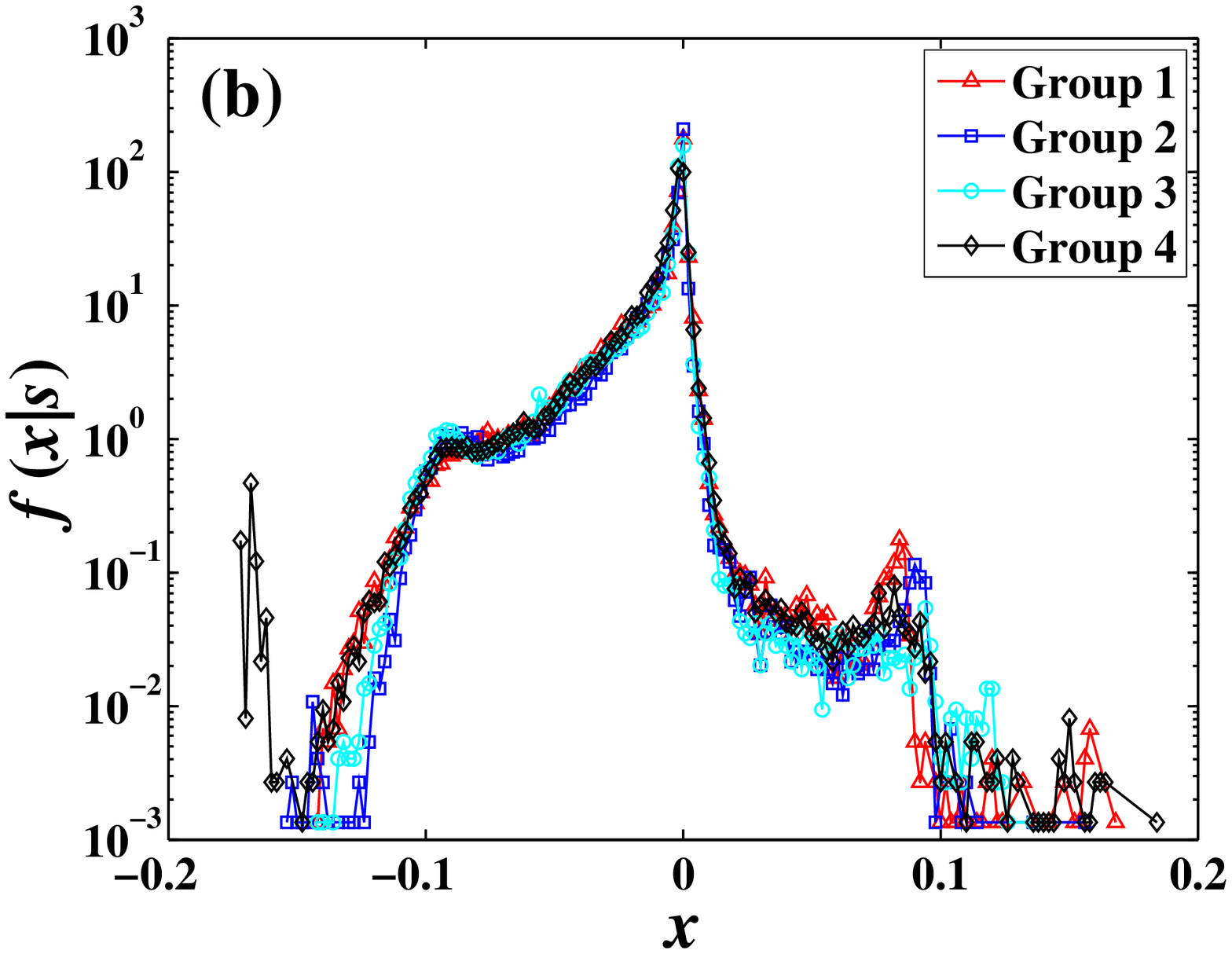}
\caption{\label{Fig:CD:spd} Conditional distribution functions on
bid-ask spread for buy orders (a) and sell orders (b) of stock
000001.}
\end{figure}

Our test is based on the idea of Mike and Farmer, who found that the
distribution of relative prices of orders is independent of the
spread \cite{Mike-Farmer-2008-JEDC}. For each order of price
$\pi(t)$ placed at time $t$, the associated spread is $s(t-1)$ right
before the order placement. The sequence of relative prices is
sorted based on the associated spreads. The data set of relative
prices is then grouped into four subsets, each of them has identical
size. For each group, we calculate the empirical conditional
distributions $f(x(t)|s(t-1))$ for buy orders and sell orders, which
are shown in Fig.~\ref{Fig:CD:spd}. We find that the probability
density functions in the four groups for each type of orders are
almost the same, independent of the bid-ask spread. The test
confirms the empirical results of Mike and Farmer
\cite{Mike-Farmer-2008-JEDC}. The result is however contrary to the
conclusion proposed by Ranaldo \cite{Ranaldo-2004-JFM}, in which the
orders are classified based on aggressiveness and the method is thus
a coarse graining of the density function.

It is interesting to note that the $f(x)$ functions corresponding to
large spread (the fourth group) have abnormal increase in the
positive tail for buy orders in Fig.~\ref{Fig:CD:spd}(a) and in the
negative tail for sell orders in Fig.~\ref{Fig:CD:spd}(b). These
phenomena can be explained partly as follows. When the spread is
large, the stock price moves fast. When the price increases, buyers
is willing to execute orders immediately so that they place orders
at the most aggressive price $\pi_{\max}$, while the sellers (say,
the $T+0$ speculators) place orders at the least aggressive price
$\pi_{\min}$.

\subsection{Conditional distribution on volatility}
\label{s2:CD:vol}

Volatility is a measure of asset risk in financial markets. We
define the volatility as the local average of the absolute returns
\begin{equation}
 v(t) = \frac{1}{N}\sum_{i=t-N+1}^{t}|\pi_m(t)-\pi_m(t-1)|
 \label{Eq:vt}
\end{equation}
where
\begin{equation}
 \pi_m(t) = [\pi_a(t)+\pi_b(t)]/2
 \label{Eq:pi:m}
\end{equation}
is the mid-price of the best ask and best bid at time $t$ and $N$ is
the number of returns $r(t)$. Here we use $N = 50$.

\begin{figure}[htb]
\centering
\includegraphics[width=6.5cm]{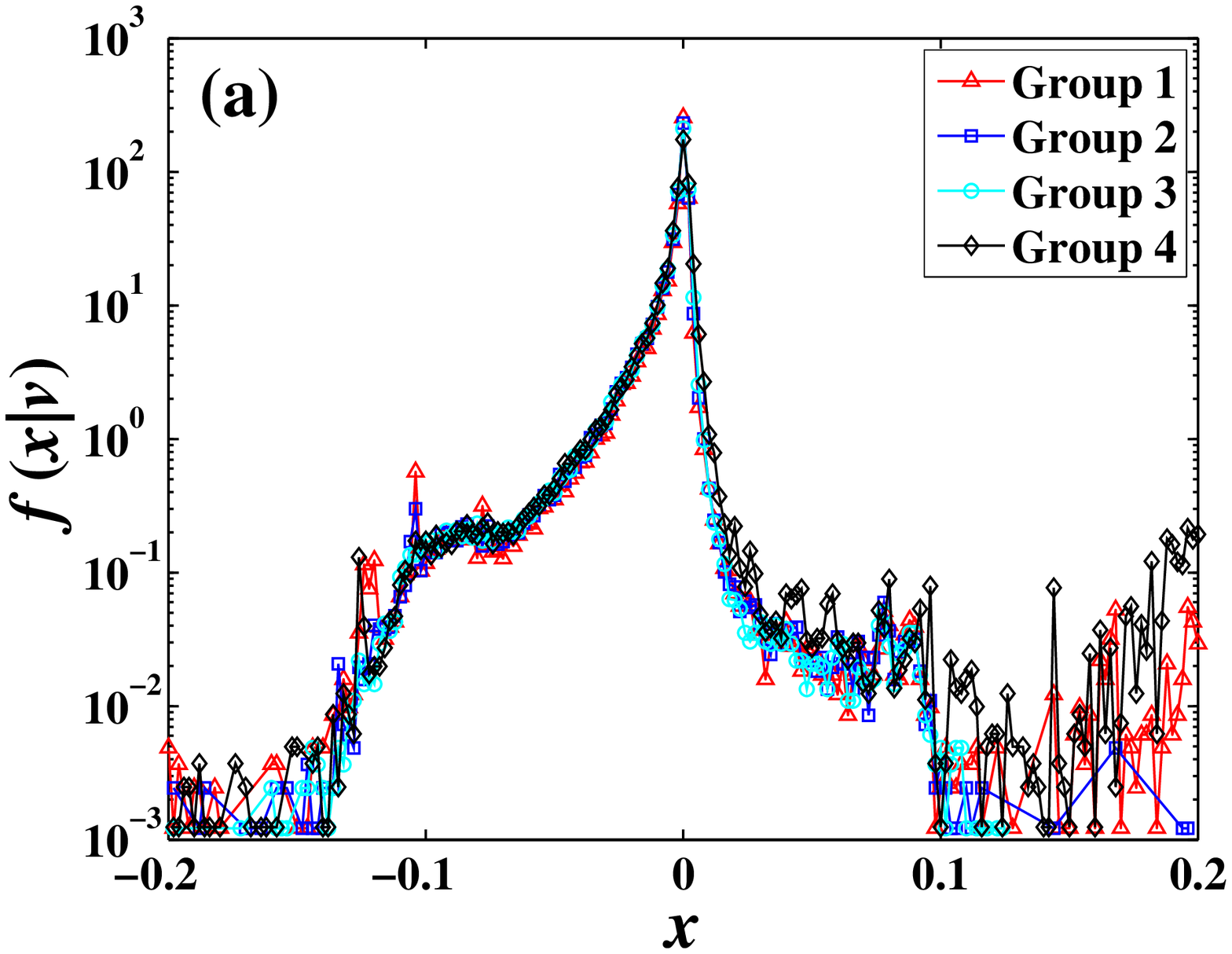}
\includegraphics[width=6.5cm]{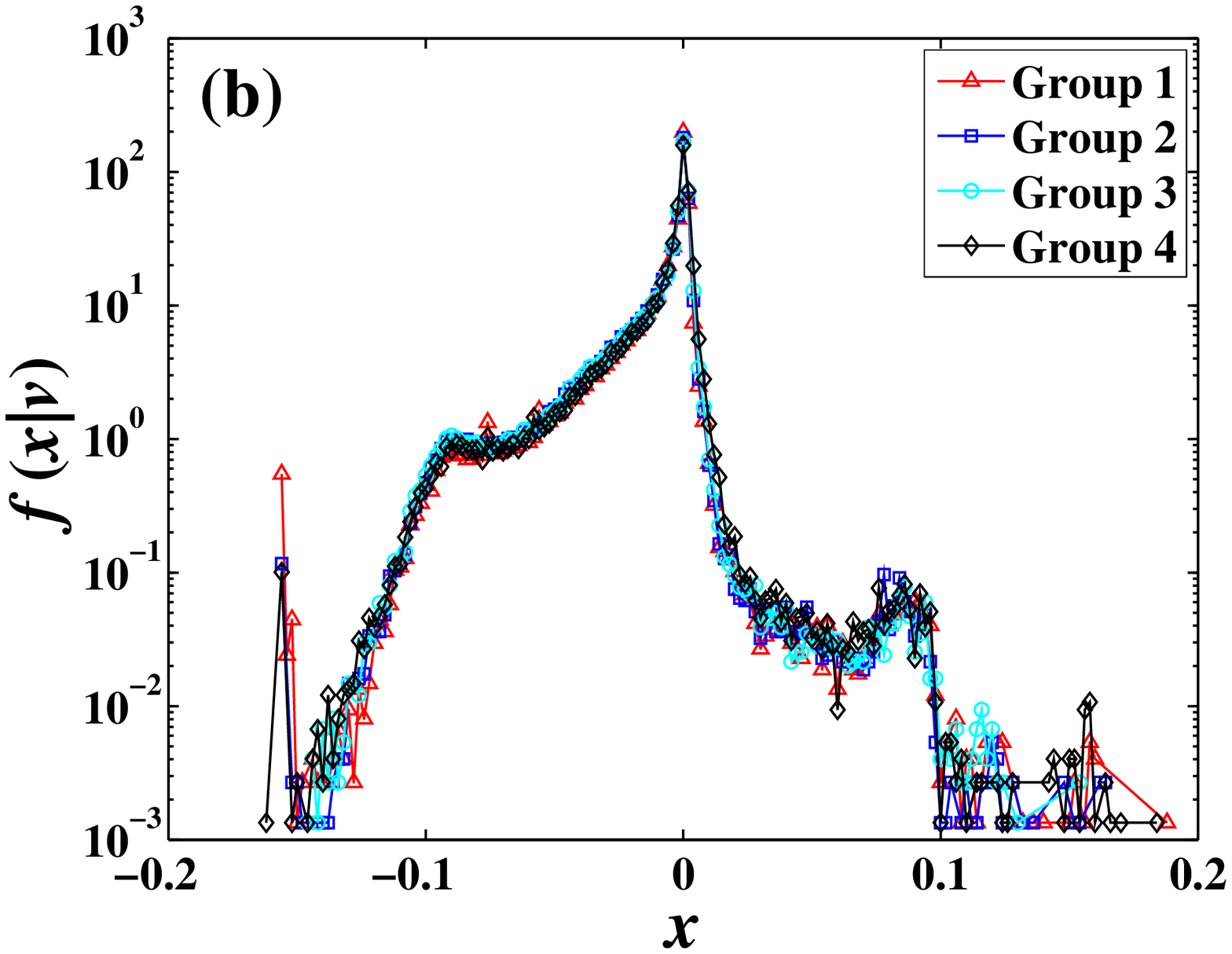}
\caption{\label{Fig:CD:vol} Conditional distribution functions on
volatility for buy orders (a) and sell orders (b) of stock 000001.}
\end{figure}

In order to investigate whether order placement is related to
volatility defined in Eq.~(\ref{Eq:vt}), we divide the whole data
into four groups with same size and increasing volatility. The
grouping procedure is similar to that used in Sec.~\ref{s2:CD:spd}.
For each group, we calculate the empirical conditional distributions
$f(x(t)|v(t-1))$ for both buy and sell orders. The results are
illustrated in Fig.~\ref{Fig:CD:vol}. We find that the probability
density functions collapse to the same curve for both buy and sell
orders, which means that order placement is almost independent of
volatility. This conclusion is consistent with the result purposed
by Lillo when he studied the limit order price of stock AZN traded
on the London Stock Exchange \cite{Lillo-2007-EPJB} but not in line
with the conclusion of Ranaldo \cite{Ranaldo-2004-JFM}.

\section{Conclusion}
\label{s1:conclusion}

We have investigated the distributions of relative prices of orders
placed in the opening call auction, cool period and continuous
double auction using ultra-high-frequency data reconstructed from
the quotes database and the trades database of 23 stocks traded on
the Shenzhen Stock Exchange within the whole year 2003. The results
for individual stocks are similar to one another, which allows us to
aggregate all 23 stocks for analysis.

The distributions in the three time periods exhibit common
properties and idiosyncratic features. During the three time
periods, the probability density function reaches the maximum at $x
= 0$, which means that a large proportion of orders are placed on
the same best price. The distributions are asymmetric between buy
orders and sell orders in each period. In addition, each
distribution is asymmetric to the same best price $x=0$ and there
are more orders placed inside the book ($x<0$). More interestingly,
all the distributions are heavily influenced by the idiosyncratic
trading rule of 10\% price fluctuation limit, which induces kinks
around $x=\pm0.1$. The distributions of large relative prices beyond
the kinks decay faster than in the bulks in exponential forms.

We have also studied the dependence of the distributions with
respect to the bid-ask spread and volatility in the period of
continuous auction, taking a very liquid stock (000001) as an
example. We found that the distributions of relative prices are
independent of the bid-ask spread and volatility.

\bigskip
{\textbf{Acknowledgments:}}

We are grateful to J. Doyne Farmer and Szabolcs Mike for
discussions. This work was partly supported by the National Natural
Science Foundation of China (Grant Nos. 70501011 and 70502007), the
Fok Ying Tong Education Foundation (Grant No. 101086), and the
Program for New Century Excellent Talents in University (Grant No.
NCET-07-0288).

\bibliography{E:/Papers/Auxiliary/Bibliography}

\begin{thebibliography}{10}
\expandafter\ifx\csname url\endcsname\relax
  \def\url#1{\texttt{#1}}\fi
\expandafter\ifx\csname urlprefix\endcsname\relax\def\urlprefix{URL }\fi

\bibitem{Cont-2001-QF}
R.~Cont, {Empirical properties of asset returns: Stylized facts and statistical
  issues}, Quant. Finance 1 (2001) 223--236.

\bibitem{Mantegna-Stanley-2000}
R.~N. Mantegna, H.~E. Stanley, {An Introduction to Econophysics: Correlations
  and Complexity in Finance}, Cambridge University Press, Cambridge, 2000.

\bibitem{Zhou-2007}
W.-X. Zhou, {A Guide to Econophysics (in Chinese)}, Shanghai University of
  Finance and Economics Press, Shanghai, 2007.

\bibitem{Cont-Bouchaud-2000-MeD}
R.~Cont, J.-P. Bouchaud, {Herd behavior and aggregate fluctuations in financial
  markets}, Macroecon. Dyn. 4 (2000) 170--196.

\bibitem{Eguiluz-Zimmermann-2000-PRL}
V.~Egu{\'i}luz, M.~Zimmermann, {Transmission of information and herd behavior:
  An application to financial markets}, Phys. Rev. Lett. 85 (2000) 5659--5662.

\bibitem{Stauffer-1998-AP}
D.~Stauffer, {Can percolation theory be applied to the stock market?}, Ann.
  Phys. 7 (1998) 529--538.

\bibitem{Stauffer-Penna-1998-PA}
D.~Stauffer, T.~J.~P. Penna, {Crossover in the Cont-Bouchaud percolation model
  for market fluctuations}, Physica A 256 (1998) 284--290.

\bibitem{DHulst-Rodgers-2000-IJTAF}
R.~D'Hulst, G.~J. Rodgers, {Exact solution of a model for crowding and
  information transmission in financial markets}, Int. J. Theoret. Appl. Fin. 3
  (2000) 609--616.

\bibitem{Xie-Wang-Quan-Yang-Hui-2002-PRE}
Y.-B. Xie, B.-H. Wang, H.-J. Quan, W.-S. Yang, P.-M. Hui, {Finite-size effect
  in the Egu{\'i}luz and Zimmermann model of herd formation and information
  transmission}, Phys. Rev. E 65 (2002) 046130.

\bibitem{Bornholdt-2001-IJMPC}
S.~Bornholdt, {Expectation bubbles in a spin model of markets: Intermittency
  from frustration across scales}, Int. J. Modern Phys. C 12 (2001) 667--674.

\bibitem{Iori-1999-IJMPC}
G.~Iori, {Avalanche dynamics and trading friction effects on stock market
  returns}, Int. J. Modern Phys. C 10 (1999) 1149--1162.

\bibitem{Foellmer-1974-JMe}
H.~F{\"o}ellmer, {Random economies with many interacting agents}, J.
  Macroeconomics 1 (1974) 51--62.

\bibitem{Chowdhury-Stauffer-1999-EPJB}
D.~Chowdhury, D.~Stauffer, {A generalized spin model of financial markets},
  Eur. Phys. J. B 8 (1999) 477--482.

\bibitem{Zhou-Sornette-2007-EPJB}
W.-X. Zhou, D.~Sornette, {Self-organizing Ising model of financial markets},
  Eur. Phys. J. B 55 (2007) 175--181.

\bibitem{Kaizoji-2000-PA}
T.~Kaizoji, {Speculative bubbles and crashes in stock markets: An
  interacting-agent model of speculative activity}, Physica A 287 (2000)
  493--506.

\bibitem{Challet-Zhang-1997-PA}
D.~Challet, Y.-C. Zhang, {Emergence of cooperation and organization in an
  evolutionary game}, Physica A 246 (1997) 407--418.

\bibitem{Challet-Marsili-Zhang-2005}
D.~Challet, M.~Marsili, Y.-C. Zhang, {Minority Games: Interacting Agents in
  Financial Markets}, Oxford University Press, Oxford, 2005.

\bibitem{Arthur-1994-AER}
W.~B. Arthur, {Inductive reasoning and bounded rationality}, Am. Econ. Rev. 84
  (1994) 406--411.

\bibitem{Challet-Marsili-Zhang-2000-PA}
D.~Challet, M.~Marsili, Y.-C. Zhang, {Modeling market mechanism with minority
  game}, Physica A 276 (2000) 284--315.

\bibitem{Jefferies-Hart-Hui-Johnson-2001-EPJB}
P.~Jefferies, M.~L. Hart, P.-M. Hui, N.~F. Johnson, {From market games to
  real-world markets}, Eur. Phys. J. B 20 (2001) 493--501.

\bibitem{Challet-Marsili-Zhang-2001-QF}
D.~Challet, M.~Marsili, Y.-C. Zhang, {From minority games to real markets},
  Quant. Finance 1 (2001) 168--176.

\bibitem{Challet-Marsili-Zhang-2001a-PA}
D.~Challet, M.~Marsili, Y.-C. Zhang, {Stylized facts of financial markets and
  market crashes in minority games}, Physica A 294 (2001) 514--524.

\bibitem{Challet-Marsili-Zhang-2001b-PA}
D.~Challet, M.~Marsili, Y.-C. Zhang, {Minority games and stylized facts},
  Physica A 299 (2001) 228--233.

\bibitem{Bak-Paczuski-Shubik-1997-PA}
P.~Bak, M.~Paczuski, M.~Shubik, {Price variations in a stock market with many
  agents}, Physica A 246 (1997) 430--453.

\bibitem{Lux-Marchesi-1999-Nature}
T.~Lux, M.~Marchesi, {Scaling and criticality in a stochastic multi-agent model
  of a financial market}, Nature 397 (1999) 498--500.

\bibitem{Mike-Farmer-2008-JEDC}
S.~Mike, J.~D. Farmer, {An empirical behavioral model of liquidity and
  volatility}, J. Econ. Dyn. Control 32 (2008) in press.

\bibitem{Maslov-2000-PA}
S.~Maslov, {Simple model of a limit order-driven market}, Physica A 278 (2000)
  571--578.

\bibitem{Daniels-Farmer-Gillemot-Iori-Smith-2003-PRL}
M.~G. Daniels, J.~D. Farmer, L.~Gillemot, G.~Iori, E.~Smith, {Quantitative
  model of price diffusion and market friction based on trading as a
  mechanistic random process}, Phys. Rev. Lett. 90 (2003) 108102.

\bibitem{Farmer-Patelli-Zovko-2005-PNAS}
J.~D. Farmer, P.~Patelli, I.~I. Zovko, {The predictive power of zero
  intelligence in financial markets}, Proc. Natl. Acad. Sci. USA 102 (2005)
  2254--2259.

\bibitem{Zovko-Farmer-2002-QF}
I.~Zovko, J.~D. Farmer, {The power of patience: A behavioural regularity in
  limit-order placement}, Quant. Finance 2 (2002) 387--392.

\bibitem{Bouchaud-Mezard-Potters-2002-QF}
J.-P. Bouchaud, M.~M{\'e}zard, M.~Potters, {Statistical properties of stock
  order books: empirical results and models}, Quant. Finance 2 (2002) 251--256.

\bibitem{Potters-Bouchaud-2003-PA}
M.~Potters, J.-P. Bouchaud, {More statistical properties of order books and
  price impact}, Physica A 324 (2003) 133--140.

\bibitem{Maskawa-2007-PA}
J.-I. Maskawa, {Correlation of coming limit price with order book in stock
  markets}, Physica A 383 (2007) 90--95.

\bibitem{Ranaldo-2004-JFM}
A.~Ranaldo, {Order aggressiveness in limit order book markets}, Journal of
  Financial Markets 7 (2004) 53--74.

\bibitem{Lillo-2007-EPJB}
F.~Lillo, {Limit order placement as an utility maximization problem and the
  origin of power law distribution of limit order prices}, Eur. Phys. J. B 55
  (2007) 453--459.

\bibitem{Gu-Chen-Zhou-2008-PA}
G.-F. Gu, W.~Chen, W.-X. Zhou, {Empirical distributions of Chinese stock
  returns at different microscopic timescales}, Physica A 387 (2008) 495--502.

\bibitem{Gu-Chen-Zhou-2007-EPJB}
G.-F. Gu, W.~Chen, W.-X. Zhou, {Quantifying bid-ask spreads in the Chinese
  stock market using limit-order book data: Intraday pattern, probability
  distribution, long memory, and multifractal nature}, Eur. Phys. J. B 57
  (2007) 81--87.

\end{thebibliography}

\end{document}